# Steganography - coding and intercepting the information from encoded pictures in the absence of any initial information

M. Kwiatkowska [I] and L. Swierczewski [II]

[I] Maria Curie-Skłodowska University in Lublin
Faculty of Mathematics, Physics and Computer Science
Pl. Marii Curie-Sklodowskiej 5
20-031 Lublin, Poland

[II] Computer Science and Automation Institute
College of Computer Science and Business Administration in Łomża
ul. Akademicka 14
18-400 Łomża, Poland

[II] Corresponding author: lswiercz@icm.edu.pl

## Abstract

The work includes implementation and extraction algorithms capabilities test, without any additional data (starting position, the number of bits used, gap between the amount of data encoded) information from encoded files (mostly images). The software is written using OpenMP standard [1], which allowed them to run on parallel computers. Performance tests were carried out on computers, Blue Gene/P [2], Blue Gene/Q [3] and the system consisting of four AMD Opteron 6272 [4]. Source code is available under GNU GPL v3 license and are available in a repository OLib [5].

## I. Introduction

Steganography is the science of determining how to conduct communication so that the presence of the message could not be detected by third parties. It differs from cryptography in a way that in cryptography existence of the message is not negated but its content remains implicit. Steganography hides the fact that any communication has been conducted.

Differences between steganography and cryptography is shown in Table 1.

|  | Cryptography | Steganography |
|---|---|---|
| Transforming information into a form incomprehensible to third parties | Yes | Yes |
| Hiding information | No | Yes |
| Key usage | Yes | Yes |
| Hiding the fact of communication | No | Yes |
| Ensuring anonymity of communicating parties | No | Yes |

| The amount of information transmitted in the communication process | Comparable to the amount of encrypted information | Much greater than the amount of encrypted information |
|---|---|---|
| Additional carrier needed | No | Yes |

TABLE 1. DIFFERENCES BETWEEN STEGANOGRAPHY AND CRYPTOGRAPHY.

## II. TRADITIONAL STEGANOGRAPHY

The use of steganography dates back to the time of Herodotus, the fifth century BC. Examples of traditional steganography can be tattooing the scalp (after the hair grew back information remains invisible). One of the best solutions of this kind applied by the Germans during World War II - microdots technique. It was based on minimazing pictures to such scale so that you can paste them into the text as a dot.

## III. DIGITAL STEGANOGRAPHY

With the development of digital technology, steganography has found a new use also in the field of science. Digital steganography bases on making subtle changes to the original medium, and therefore gives a much more possibilities than traditional steganography.

The carrier of concealed information can be virtually any file (one that can be modified without having to worry about damage to its internal structures). However, the most commonly used are multimedia files - these are relatively large in size and difficult to capture the modification of original file.

Digital steganography depending on the type of operation can be divided into the following categories:

- Substitutional
- Transformational
- Spectrum modification
- Spectrum spread
- Distortional
- Statistical
- Carrier generation

Another field of use for steganography is to communicate using VoIP technology. The Polish jargon adopted using in this case the word „steganphony". The first such solution had been proposed by two Polish scientists Wojciech Mazurczyk and Krzysztof Szczypiorski in 2008 at a conference in Mexico [11]. It was based on in transferring the hidden content in the delayed packets, which according to a standard communication protocol (operating in real time) are omitted.

## IV. LSB - LEAST SIGNIFICANT BIT - ONE OF THE METHODS OF SUBSTITUTION

The article will summarize often used method of substitution - LSB. Mostly it uses the least significant bits to record information. These bits often carry only noise and are insignificant from the point of view images for example. The principle of LSB operation for one and two least significant bits are shown in Figure 1 and Figure 2.

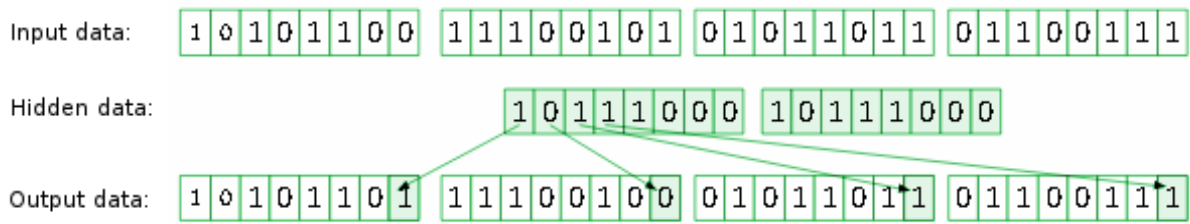

FIGURE 1. APPLYING THE LSB USING ONE LEAST SIGNIFICANT BIT.

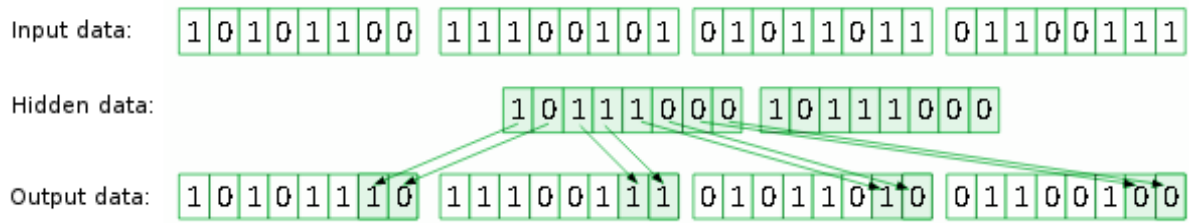

FIGURE 2. APPLICATION OF THE LSB METHOD USING THE TWO LEAST SIGNIFICANT BITS.

## V. IMPLEMENTATION

Source code had been implemented in pure ANSI C. Sample code for function encrypting text using least significant bits of the image you can see in Listing 1.

```
int crypt_space_1bit(char *buffer, char *open_text, long int start_position, long int open_text_size, long int space_size)
{
    long int i;
    for(i=0; i < (open_text_size * 8); i++)
    {
        if( check_bit(open_text[i/8], i%8) )
        {
            set_bit(buffer[start_position+i*space_size], 0);
        }
        else
        {
            clear_bit(buffer[start_position+i*space_size], 0);
        }
    }
}
```
LISTING 1 THE FUNCTION ENCRYPTING THE DATA IN THE IMAGE USING ONE LEAST SIGNIFICANT BIT.

We are passing pointers to two array to the function - the *buffer* array (image data) and *open_text* array (array of information to be encoded). In addition, the code uses macros check_bit, set_bit and clear_bit. Their definition is shown in Listing 2.

```
# define check_bit(var,pos) ((var) & (1LL<<(pos)))
# define set_bit(var,pos) ((var) |= (1LL<<(pos)))
# define clear_bit(var,pos) ((var) &= ~(1LL<<(pos)))
```

LISTING 2. MACROS DEFINING OPERATIONS ON BITS.

The most important, in the case of this article, however, is a function used to extract encrypted text without knowledge of the initial starting value, or even the length of ciphertext. It has been shown in Listing 3.

```
long int breaker_standard_1bit_unknown_size_openmp(char *buffer, long int buffer_size, long int size_down, long int size_upper, result_structure *result, int number_of_threads)
{
        char *open_text;

        long int start_position;

        long int index_result;
        index_result = 0;

        long int i;
        long int j;
        j = size_down;

        #pragma omp parallel for shared(buffer, buffer_size, size_down, size_upper, result, index_result) private(i, j, start_position, open_text) num_threads(number_of_threads)
        for(i=0; i < (buffer_size - j); i++)
        {
                open_text = (char*) malloc( (size_upper+1) * sizeof(char));

                start_position = i;

                for(j=size_down; j <= size_upper; j++)
                {
                        encrypt_standard_1bit(buffer, open_text, start_position, j);

                        if ( ascii_veryfi(open_text, j) == 1 )
                        {
                                open_text[j] = '\0';

                                #pragma omp critical
                                {
                                        result[index_result].start_position = i;
                                        strcpy(result[index_result].text, open_text);
                                        index_result++;
                                }
                        }
                }
```

```
            }
            free(open_text);
        }
    return index_result;
}
```
LISTING 3 FUNCTION TO EXTRACT INFORMATION FROM THE LEAST SIGNIFICANT BITS WITHOUT THE INITIAL INFORMATION.

This function searches the least significant bits in the *buffer* array of size of *buffer_size* and searches all the possible encrypted patterns of lengths from *size_down* to *size_upper*. The result is inserted into the result array. The code uses so-called pragma derived from the OpenMP standard. This pragma aims to parallelize the main loop. Also worth mentioning is separate critical section - it takes care of the correct addition of blocks to the *results* array. Only one thread can access the critical section at a time. Also a function has been used:

   int ascii_verify(pointer, size);

which verifies whether the starting substring at the *pointer* of length *size* is the correct text stored using ASCII code.

## VI. RESULTS

Performance results obtained when searching for encoded text inside graphic file with a resolution of 1600x1200 is shown in Table 2. A similar table for resolution 4096x4096 is presented in Table 3. Additionally, obtained through parallel programming (OpenMP), speedup on IBM Blue Gene/Q and AMD Opteron 6272 is shown in Figure 3.

| Number of threads | Execution time | | |
|---|---|---|---|
| | IBM Blue Gene/P | IBM Blue Gene/Q | AMD Opteron 6272 |
| 1 | 560.47 s | 193.11 s | 180.00 s |
| 2 | 282.78 s | 97.71 s | 92.00 s |
| 4 | 149.23 s | 48.38 s | 52.00 s |
| 6 | - | 32.28 s | 34.00 s |
| 8 | - | 24.27 s | 25.00 s |
| 12 | - | 16.29 s | 21.00 s |
| 16 | - | 12.32 s | 16.00 s |
| 32 | - | 6.77 s | 8.00 s |
| 48 | - | 5.64 s | 6.00 s |
| 64 | - | 7.60 s | 5.00 s |

TABLE 2. PERFORMANCE RESULTS OBTAINED DURING SCAN OF THE IMAGE WITH A RESOLUTION OF 1600X1200 (SEARCHING ONE LEAST SIGNIFICANT BIT USING LENGTH OF THE SEARCH PHRASE IN THE RANGE OF 10 TO 25).

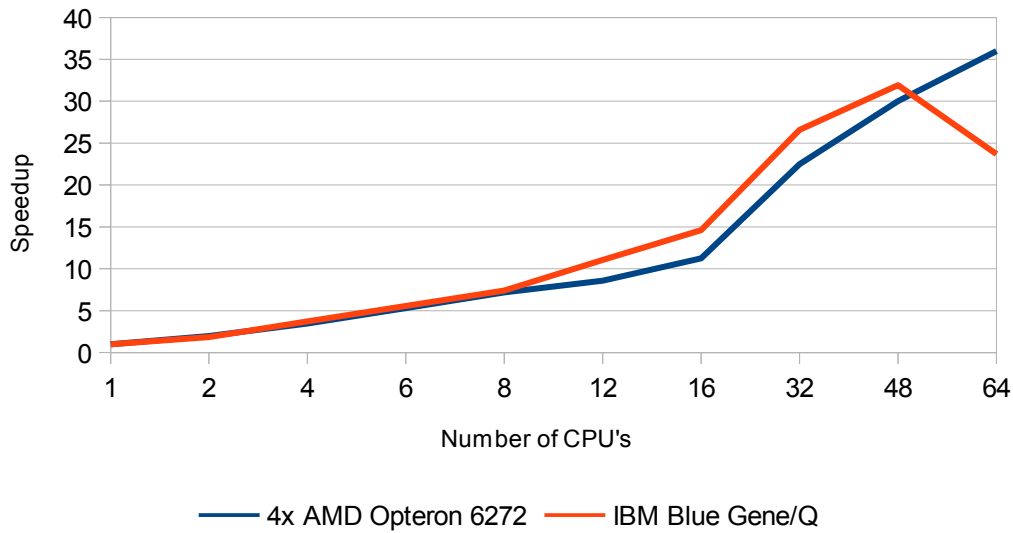

FIGURE 3. THE SPEEDUP OBTAINED ON PLATFORMS AMD OPTERON 6272 AND IBM BLUE GENE/Q WHILE SEARCHING AN IMAGE WITH A RESOLUTION OF 1600x1200 (SEARCH ONLY ONE LEAST SIGNIFICANT BIT OF THE LENGTH OF THE TEXT TO SEARCH IN THE RANGE FROM 10 TO 25).

|   | Execution time | | |
|---|---|---|---|
| Number of threads | IBM Blue Gene/P | IBM Blue Gene/Q | AMD Opteron 6272 |
| 1 | 4769.00 s | 1309.42 s | 1282.08 s |
| 2 | 2482.00 s | 651.12 s | 652.00 s |
| 4 | 1311.00 s | 308.89 s | 401.00 s |
| 6 | - | 201.24 s | 265.00 s |
| 8 | - | 147.00 s | 200.00 s |
| 12 | - | 97.25 s | 132.00 s |
| 16 | - | 74.19 s | 107.00 s |
| 32 | - | 36.90 s | 59.00 s |
| 48 | - | 25.76 s | 43.00 s |
| 64 | - | 22.49 s | 35.00 s |

TABLE 3. PERFORMANCE RESULTS OBTAINED DURING SCAN OF THE IMAGE WITH A RESOLUTION OF 4096x4096 (SEARCHING ONE LEAST SIGNIFICANT BIT USING LENGTH OF THE SEARCH PHRASE IN THE RANGE OF 10 TO 25).

In the case of a platform consisting of four AMD Opteron 6272 maximum speedup is achieved by using 64 CPU and it reached 36.00. Using computational units used in Blue Gene/Q the speedup reached to 31.915. But it was not obtained, as might be expected, with a maximum (64) number of processors, but only 48. This may be due to the fact that one Blue Gene/Q CPU has only 16 physical cores that implement the execution of 64 threads.

It can also be noted that increasing the resolution of the analyzed image from 1600x1200 to 4096x4096 did not increase the runtime of the algorithm even tenfold - for applications executed sequentially on the Blue Gene/Q time increased from 193.11 seconds to 1309.42 seconds (about

6.78 times) and for the Blue Gene/P changed from 560.47 seconds to 4769.00 (about 8.50 times). Analysis has been performed for the speedup depending on the size of the hidden string searched. The results are shown in Table 3 (for AMD Opteron 6272) and in Table 4 (for Blue Gene/Q). Presentation of speedup obtained is presented in Figure 4. Additionally, Figure 5 shows the difference in the runtime of the sequential program execution between two platforms (AMD Opteron 6272 - IBM Blue Gene/Q) .

| Number of CPU's | *Down_Size = 10; Upper_Size =* | | | | | | | | | |
|---|---|---|---|---|---|---|---|---|---|---|
| | 15 | 20 | 25 | 30 | 35 | 40 | 45 | 50 | 55 | 60 |
| 1 | 348.47 | 757.28 | 1282.40 | 1919.38 | 2668.21 | 3524.63 | 4501.62 | 5631.80 | 6799.42 | 8122.23 |
| 64 | 13.00 | 24.00 | 34.00 | 51.00 | 72.00 | 99.00 | 125.00 | 158.00 | 193.00 | 192.00 |
| Speedup | 26.805 | 31.553 | 37.718 | 37.635 | 37.064 | 35.602 | 36.013 | 35.664 | 35.230 | 42.303 |

TABLE 4. THE RESULTS OF THE ANALYSIS OF PERFORMANCE FOR DIFFERENT SIZES OF SEARCHED HIDDEN STRING [DOWN_SIZE; UPPER_SIZE] (THE RANGES OF [10, 15] TO [10,60]) AND PLATFORM BASED ON AMD OPTERON 6272.

| Number of CPU's | *Down_Size = 10; Upper_Size =* | | | | | | | | | |
|---|---|---|---|---|---|---|---|---|---|---|
| | 15 | 20 | 25 | 30 | 35 | 40 | 45 | 50 | 55 | 60 |
| 1 | 353.18 | 773.93 | 1295.46 | 1949.57 | 2692.76 | 3562.74 | 4545.64 | 5646.70 | 6863.10 | 8222.39 |
| 64 | 9.43 | 15.64 | 22.00 | 31.90 | 43.01 | 55.16 | 67.39 | 81.99 | 98.64 | 116.93 |
| Speedup | 37.452 | 49.484 | 58.885 | 61.115 | 62.608 | **64.589** | **67.452** | **68.871** | **69.577** | **70.319** |

TABLE 5. THE RESULTS OF THE ANALYSIS OF PERFORMANCE FOR DIFFERENT SIZES OF SEARCHED HIDDEN STRING [DOWN_SIZE; UPPER_SIZE] (THE RANGES OF [10, 15] TO [10,60]) AND PLATFORM BASED ON PROCESSORS INSTALLED IN THE BLUE GENE/Q.

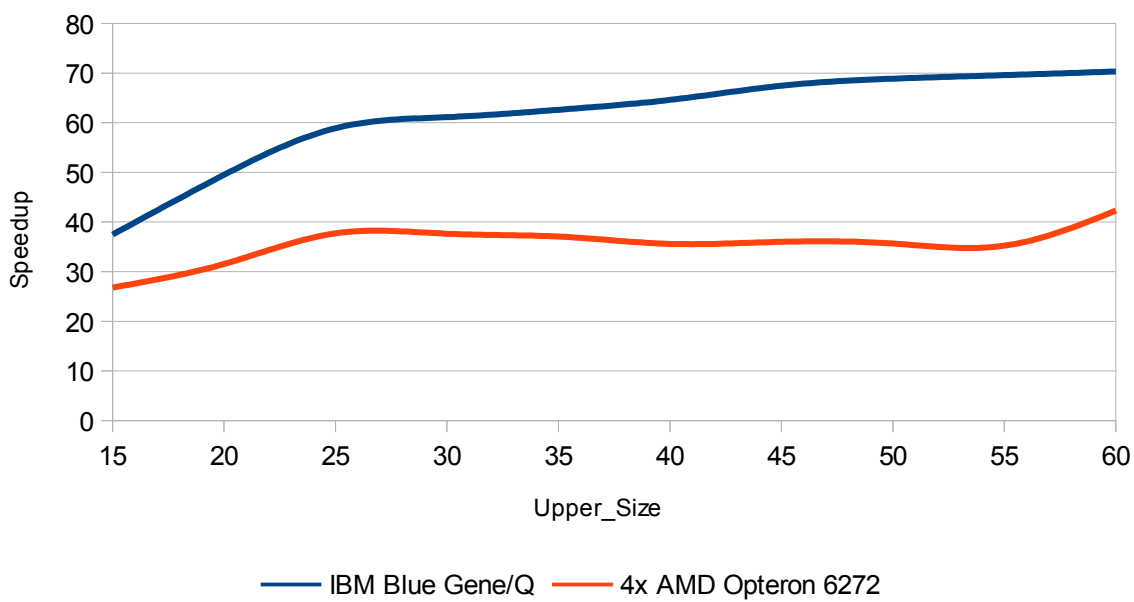

FIGURE 4. GRAPH SHOWING THE RESULTING SPEEDUP ON AMD OPTERON 6272 AND IBM BLUE GENE/Q IN THE ANALYSIS OF PERFORMANCE FOR DIFFERENT SIZES OF SEARCHED HIDDEN STRING.

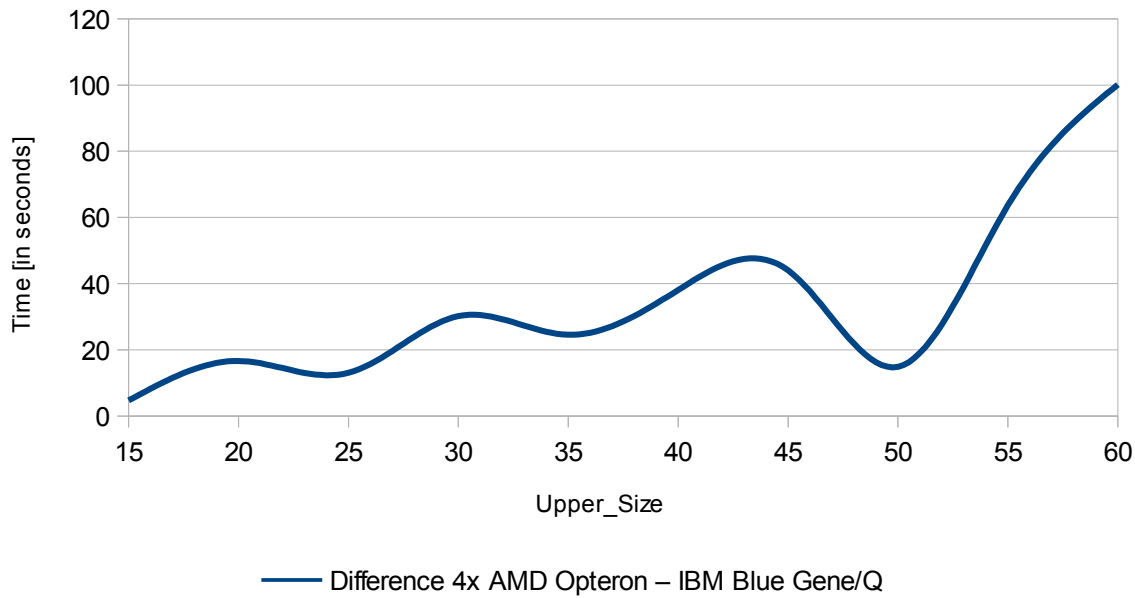

FIGURE 5. GRAPH SHOWING THE TIME DIFFERENCE OBTAINED ON A PLATFORM CONSISTING OF THE AMD OPTERON 6272 AND OBTAINED ON THE BLUE GENE/Q IN THE ANALYSIS OF PERFORMANCE FOR DIFFERENT SIZES OF SEARCHED HIDDEN STRING.

The strangest fact can be seen in the analysis of speedup presented in Table 5 According to the measurements on 64 processors for searching string of a length of 10 to 60 characters, maximum achieved speedup of 70.319, which is impossible and must be the result of the time measurment error. Problematic data in the table indicated in red. This anomaly is no longer present on the computer with four AMD Opteron 6272. However, in this case, the maximum speedup was only 42.303 and was obtained while searching a string with the length in the range [10, 60].

While in terms of speedup, the platform for the Blue Gene/Q is better but the lower sequential execution runtime of the program is always achieved on a computational node with AMD Opteron 6272 - it is confirmed in Figure 5. Differences in the runtime between these two systems, however, are relatively small and do not exceed *2%*.

### VII. CONCLUSIONS AND CAPABILITIES OF WORK DEVELOPMENT

Using the abilities of parallel programming, one can very effectively deal with data processing in the field of steganography. LSB technique is used in most commercial software.

The work can be treated as a base for discussion, because it is very difficult to find a practical application of implemented steganography methods in the modern world. In the so-called *„pure steganography"* strenght of encryption is mainly based on lack of knowledge about the techniques used by the individual encrypting the information [7]. So you can not publish this type of algorithms as Open Source. Such an approach, however, does not meet Kerckhoffs principle [10] stating, that a cryptographic system should be secured even if all the details about how it works (except the key) are known, and it is not recommended .

The presented work results concern the simplest variant in which the string is interpreted as ASCII code. Of course you can come up with your own, much more complicated system for the representation of characters.

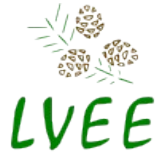

FIGURE 6. LOGO OF THE CONFERENCE LVEE WITH HIDDEN TEXT „*LVEE – THE BEST CONFERENCE*".

**ACKNOWLEDGMENT**

Interdisciplinary Centre for Mathematical and Computational Modeling (ICM), Warsaw University, Poland is acknowledged for providing the computer facilities under the Grant No. G55-11.

**NOTE**

The thesis presented on Winter LVEE 2014 Conference in Minsk (Belarus).

# Steganografia – kodowanie oraz przechwytywanie informacji z zakodowanych obrazów przy braku jakichkolwiek informacji początkowych

M. Kwiatkowska [I] i L. Swierczewski [II]

[I] Uniwersytet Marii Curie-Skłodowskiej w Lublinie
Wydział Matematyki, Fizyki i Informatyki
Pl. Marii Curie-Sklodowskiej 5
20-031 Lublin, Polska

[II] Państwowa Wyższa Szkoła Informatyki i Przedsiębiorczości w Łomży
ul. Akademicka 14
18-400 Łomża, Polska

[II] Adres do korespondencji: lswiercz@icm.edu.pl

## Streszczenie

Praca obejmuje implementację oraz test możliwości algorytmów wyciągających bez jakichkolwiek dodatkowych danych (pozycja startowa, ilość wykorzystanych bitów, wykorzystany odstęp między zakodowanymi porcjami danych) informacje z zakodowanych plików (głównie obrazów). Oprogramowanie napisano z wykorzystaniem standardu OpenMP [1], co umożliwiło uruchomienie ich na komputerach równoległych. Testy wydajnościowe przeprowadzono na komputerach, Blue Gene/P [2], Blue Gene/Q [3] oraz systemie złożonym z czterech procesorów AMD Opteron 6272 [4]. Kody źródłowe są dostępne na postawie licencji GNU GPL v3 i zostały udostępnione w repozytorium OLib [5].

**Słowa kluczowe:** *Steganografia, LSB, Blue Gene, OLib*